\documentstyle[12pt]{article}
\begin{document}
\begin{titlepage}
\begin{flushright}
Zurich University ZU-TH 5/97\\
\end{flushright}
\vfill
\begin{center}
{\large\bf Achromatic variability
in the BL Lac Object PKS 2155-304.\\
 A case for microlensing?}
\vskip 0.5cm
A.~Treves$^{1}$,
F. Rovetti$^{1}$ ,
Ph.~Jetzer$^{2}$
and C.M. Urry$^{3}$
\vskip 0.5cm
$^1$
International School for Advanced Studies, S.I.S.S.A.,
        Strada Costiera 11, I-34014 Trieste, Italy\\
$^2$
Paul Scherrer Institute, Laboratory for Astrophysics, CH-5232 Villigen PSI, and
Institute of Theoretical Physics, University of Zurich, Winterthurerstrasse
190, CH-8057 Zurich, Switzerland\\
$^3$
Space Telescope Science Institute, 3700 San Martin Drive, 
Baltimore, MD 21218, USA 

\vfill
\end{center} 
 
\begin{abstract}
\noindent
PKS 2155-304 is the only BL Lac object for which well-sampled
multiwavelength light curves resolve the intraday variability at UV
and X-ray wavelengths. In particular we focus on the multifrequency
campaign of November 1991, which showed a rather exceptional behaviour
of the source exhibiting substantial achromatic variability from optical 
to X-ray wavelengths. We suggest a scenario for this unique
event, where the variability is due 
to microlensing. The relativistic motion of the source relative to
the lens is taken into account. The lenses are proposed to be solar-mass
stars in an intervening dwarf galaxy. The a priori probability of detecting
a microlensing event, however, was only few percent, small but not 
negligible.
\end{abstract}
\vfill
\begin{flushleft}
Talk presented by A. Treves at the Girona meeting (Spain, September 1996).
To appear in the proceedings,
which will be published in Astrophysics and Space Science.
\end{flushleft}
\end{titlepage}
\newpage 
\baselineskip=21pt
 
\section{Introduction}
BL Lac objects are Active Galactic Nuclei (AGN) in which most of the observed
radiation likely comes from plasma moving relativistically toward the observer
(Blandford \& Rees 1978; Urry \& Padovani 1995).
This scenario explains characteristic blazar properties
like rapid variability, high luminosities, and high polarization
(Stein et al. 1976, Angel \& Stockman 1980), as well as (to some extent)
the low equivalent widths by which BL Lacs are defined
(Morris et al. 1991, Stickel et al. 1991).
 
The most rapid variability has been resolved only for a handful
of blazars that are bright enough or located in advantageous parts of the
sky. For example, rapid intraday variability at radio and optical wavelengths
has been observed in the BL Lac object PKS~0716+714 in part because of its
high ecliptic latitude and therefore easy accessibility to dedicated
radio and optical telescopes (Quirrenbach et al. 1989).
At shorter wavelengths, very
few blazars are bright enough to monitor with comparable temporal resolution.
Only one or two can normally be observed with IUE at sub-one-hour integration
times, and the brightest of these is PKS~2155-304.
 
Throughout November 1991, PKS~2155-304 was monitored daily with IUE, and
for $\sim4.5$ days in the middle of the month was observed nearly continuously
with 90-minute resolution (see fig.3 of 
Urry et al. 1993). At roughly the same time,
$\sim3.5$ days of continuous Rosat observations were obtained (Brinkmann
et al. 1994). The optical, UV, and X-ray light curves show strongly correlated
variations with $\sim 10$-30\% amplitude over times scales of half a day to
a day, 
with no dependence of amplitude on wavelength (Edelson et al. 1995).
This achromatic variability is a rather unique behaviour
of the source during the observation under consideration and
contrasts significantly, for instance, with the mode of variation detected
during another long exposure performed in May 1994 (Urry et al. 1997).
In terms of spectral shape, the optical through X-ray continuum from
PKS~2155-304 can be understood as synchrotron emission from
highly relativistic electrons (Urry \& Mushotzky 1982),
but achromatic variability of the kind observed
in November 1991 is very difficult to explain as variations in the physical
state of the synchrotron emitting plasma (Celotti et al. 1991).
 
In contrast, if the size of the emitting region is not a strong function of
wavelength, at least for optical through X-ray wavelengths, then
gravitational microlensing can cause achromatic variations,
since the magnification itself is wavelength independent.
For plausible lens parameters, intraday variability would result only if
the relative source-lens velocity were relativistic but this occurs naturally
in blazars assuming they have aligned relativistic jets.
It has in fact been suggested that the BL Lac phenomenon may be
explained by gravitational microlensing of Optically Violently Variable
(OVV) quasars by foreground galaxies, wherein the low equivalent widths of
BL Lacs relative to OVVs result from the compact continuum source
being magnified relative to the extended emission line gas
(Vietri \& Ostriker 1983). Whether this microlensing scenario
works for BL Lacs as a class is unclear (Urry \& Padovani 1995) ---
but it may be important for individual objects (e.g., Stickel et al. 1993,
Stocke, Wurtz \& Perlman 1995).
 
In this paper we consider gravitational microlensing of a background
relativistic source as a possible
explanation for the rapid achromatic variability observed in 
 PKS~2155-304. The effect of microlensing on source intensity
has already been studied in detail for quasars (Kayser et al. 1986,
Schneider \& Weiss 1987), and Krishna \& Subramanian (1991) have developed
the formalism for the case of relativistic source-lens velocities.
Here, we review the basic equations, discuss the location of PKS~2155-304
and intervening matter, and make a quantitative assessment of
whether microlensing can explain variability in this, the blazar with the
best multiwavelength light curves.
 
\section{Basic Lensing Theory}
 
We consider the relevant equations for lensing, evaluating them numerically for
the particular case of PKS~2155-304.
Let $D_S$ be the source-distance, $D_L$ the lens-distance and $D_{LS}$
the lens-source distance. The redshift of PKS~2155-304 is
$z=0.116$ (Falomo et al. 1993) and we adopt $D_S= 600$~Mpc
(corresponding to $H_0 = 60$~km/s/Mpc, $q_0=0.5$).
The two most likely locations for the lens are within the host
galaxy of the BL Lac itself or at the intervening redshift, $z\sim0.059$,
at which strong Ly-$\alpha$ absorption has been seen in the far-UV spectrum of
PKS~2155-304 (Bruhweiler et al. 1993).
 
\subsection{Microlensing in a Central Stellar Cluster}
 
As lenses we first consider stars in a stellar cluster which we
suppose to be located at the center of the BL Lac host galaxy.
We assume the cluster consists of $N=10^7$ stars of mass $M=1 ~ M_{\odot}$
within a radius $R_L=1$~pc.
The radius of the Einstein ring for any one star is given by the classical
equation
\begin{equation} \label{e1}
\xi=\sqrt{\frac{4GM}{c^2} \frac{D_{L}D_{LS}}{D_S}}~.
\end{equation}
Taking $D_{LS} \sim R_L = 1$~pc
and $D_L \simeq D_S$, one gets
$\xi=1.4 \times 10^{12}$~cm, or roughly $2\times10^{-10}$~arcsec.
The optical depth to lensing is
\begin{equation} \label{e2}
\tau \simeq n \sigma R_L ,
\end{equation}
where $n$ is the density of
cluster stars. In the rough approximation of homogeneity,
\begin{equation} \label{e3}
n=\frac{N}{\frac{4}{3} \pi R_L^3}
\end{equation}
and the lensing cross section is
\begin{equation} \label{e3b}
\sigma = \pi \xi^2~.
\end{equation}
For the central stellar cluster we are considering,
$\tau \sim 1.5 \times 10^{-6}$.
 
Now we further
suppose that the source moves with a large tranverse velocity $v_S$ in the
 source plane,
large enough that it dominates over stellar velocities in the lens.
One actually measures the angular velocity of the source, $\omega$,
converting to the transverse velocity at the source via $v_S = D_S \omega$.
Here, however, the relevant velocity is the transverse velocity at the
lens, $v_L = D_L \omega = \frac{D_L}{D_S} v_S$.
The effective area that can activate the gravitational focusing in the lens
plane is therefore
\begin{equation} \label{e4}
\Sigma_L \sim 2\: \xi \frac{D_L}{D_S} \:v_S\: t~.
\end{equation}
The time scale $t$ for the occurrence of a lensing episode can be expressed
through the condition
\begin{equation} \label{e5}
\Sigma_L \: n \:R_L=1~;
\end{equation}
thus from Eqns. (3), (5), and (6), we have
\begin{equation} \label{e6}
t=\frac{2 \pi R_L^2}{3 v_S N \xi} \frac{D_S}{D_L}~.
\end{equation}
The typical duration of a microlensing episode,
valid for small redshift of the source as in this case, is
\begin{equation} \label{e7}
T=\frac{\xi}{v_S} \frac{D_S}{D_L}~.
\end{equation}
(The correct expression in general is given by Eq. 5 in
Krishna \& Subramanian 1991.)
Since we have taken as a typical length scale for the focusing
the Einstein radius, the corresponding amplification factor
is $\sim30$\% (e.g., Schneider et al. 1992).
 
Eqn. (\ref{e7})
is valid in the approximation of a point-like source,
i.e., for source radius
\begin{equation}
R_S < \xi \frac{D_S}{D_L}~. \label{e8}
\end{equation}
When this approximation is not valid one should follow the treatment discussed
by Schneider and Weiss (1987).
 
Since we are considering a BL Lac object, the apparent source velocity
at us or at the lens is likely relativistic (Vermeulen \& Cohen 1994)
and we adopt a bulk Lorentz factor $\gamma=5$.
Moreover, we assume the viewing angle to be small, with a representative
value $\theta=10^{\circ}$, which results in an apparent transverse velocity
$v_S=5 c$.
 
For this choice of parameters Eqns. (\ref{e6}) and (\ref{e7}) yield:
$t \sim 110$~days, $T \sim 9$~s, which is far too rapid
for the observed variations. This remains true for other plausible
locations of stellar clusters in the host galaxy and so we turn to lenses
not associated with the host galaxy.
 
\subsection{Microlensing by an Intervening Dwarf Galaxy}
 
In the particular case of PKS~2155-304, there is a high probability
that stars could be located along the line-of-sight, approximately halfway
to the BL Lac object. This is because ultraviolet spectra of
PKS~2155-304 exhibit Ly-$\alpha$ absorption features at
$z=0.057$, 0.059, and 0.060. The $z=0.059$ system is particularly strong,
with equivalent width $W_\lambda\sim 1$~\AA\
(Maraschi et al. 1988, Allen et al. 1993,
Bruhweiler et al. 1993, Appenzeller et al. 1995).
The nature of the absorber is unknown since
no associated metal lines have yet been detected, and
because the Ly-$\alpha$ line is unresolved,
the hydrogen column density is indeterminate.
Optical images with a seeing $\leq 1$~arcsec indicate no foreground galaxies
brighter than $m_V=19$ 
within 7~arcmin of the source (Falomo et al. 1993).
If the absorber is a galaxy and not a hydrogen cloud, it must be a
dwarf galaxy very close to the line of sight to PKS~2155-304.
 
Following this scenario, the lens distance is
$D_L= 300~ {\rm Mpc} \simeq D_{LS}$, the
lens radius is $R_L= 1~ {\rm kpc}$, and
we assume the central part of the dwarf galaxy contains
$N=10^9$ stars of mass 1~$M_{\odot}$.
Using Eqns. (1), (\ref{e6}), and (\ref{e7}),
we get (with $M=1~M_{\odot}$)
$\xi = 5.4 \times 10^{-3}~ {\rm pc} = 1.66 \times 10^{16}~ {\rm cm}$,
$t=185$~d, and $T=2.6$~d. The duration is the right order of magnitude,
especially given the crude approximations used to derive Eqns.
\ref{e6} and \ref{e7}.
 
\section{Discussion}
 
BL Lacs are likely characterized by the relativistic bulk motion of their
emission regions. As is apparent from Eqns. (\ref{e6}) and (\ref{e7}),
this implies a
drastic reduction of the recurrence time and of the duration of the lensing
episodes relative to conventional microlensing of quasars. Indeed, the time
scales for microlensing considered here range from a few seconds for a
central stellar cluster in the BL Lac host galaxy to a few days for stars
in an intervening galaxy.
 
In practice, an episode with the characteristics corresponding to lensing
by a central stellar cluster would be rather hard to detect
because of its short duration, and it certainly does not correspond to the
variations seen in November 1991. Furthermore,
the condition on the source size (Eqn. \ref{e8}) implies an unreasonably
small size of the optical/UV/X-ray emitting region in PKS~2155-304,
$R_S < 10^{12}$~cm. 
 
In contrast, the time scale for microlensing
by a stellar cluster in a galaxy at $z=0.059$, the redshift of a known
Ly-$\alpha$ absorption system along the line-of-sight to PKS~2155-304,
is the right order of magnitude for the observed November 1991 light curve.
In this case, also, Eqn. \ref{e8} implies a reasonable limit
on the size of the optical/UV/X-ray emitting region in PKS~2155-304,
$R_S < 3 \times 10^{16}$~cm, easily commensurate with our understanding of
the emission mechanism in this BL Lac object.

As discussed by Urry et al. (1993) the flux variation is 
essentially achromatic. On this regard see also Brinkmann et al. (1994,
figs. 3 and 6) for the X-ray observations 
and Edelson et al. (1995, fig. 2b) for a comparison of optical, UV and
X-ray light curves.
The overall duration of the event is $\sim~4$ days. The spiky structure
apparent in the light curves may be related to the caustic web as discussed
by Schneider \& Weiss (1987) and Krishna \& Subramanian (1991) and/or
to a structure of the relativistic blobs. Moreover, one should consider
that the source is certainly variable independently of microlensing
and the light curve could represent the superposition of the two
processes. 

One concern is that microlensing events should be quite rare,
raising a question as to the probability of our having observed
the phenomenon. Specifically, with a recurrence period of 185~days,
a campaign lasting $\sim5$~days
(as in November 1991) should detect microlensing events only a few percent
of the time, a very small though non-negligible fraction.
A slightly longer monitoring campaign in May 1994
did not reveal similar low-amplitude, achromatic variability --- indeed
the wavelength dependence of the variability was quite dramatic
(Urry et al. 1997) --- but we
are clearly dealing with small-number statistics. (These
two sets of observations represent the only data with which
rapid UV/X-ray variations could have been resolved.)
 
Another problem for establishing the importance of gravitational
microlensing in PKS~2155-304 is,
as we already noticed, that different variability mechanisms are
certainly at work, as demonstrated by the contrast between the achromatic
variations in November 1991 and the strong spectral variability seen in
May 1994. Thus the possibility of confusion among different kinds of
variations exists.
 
Nevertheless, we have demonstrated that microlensing by stars
in a dwarf galaxy at the known
redshift of Ly-$\alpha$ absorption toward PKS~2155-304 is a promising scenario
for explaining its rapid, correlated, achromatic, optical/UV/X-ray variability.
The clearest proof of the validity of this hypothesis
would be the detection of the dwarf galaxy itself; however,
given the brightness of the BL Lac nucleus,
this will be difficult even with HST.
 
Whether microlensing is an important source of variability
in other BL Lacs remains to be seen. Two likely prerequisites,
following from our discussion of PKS~2155-304, are that the observed
multiwavelength variability be at least roughly achromatic
(well-correlated with lag smaller than a few hours) and that
there is some evidence for intervening matter along the line of sight.\\
 
\noindent{\bf Acknowledgements}
 
C. M.U. acknowledges support from NASA Grant NAG5-1034.


\begin{thebibliography}{}
 
\bibitem{ } Allen, R. G., Smith, P. S., Angel, J. R. P., Miller, B. W.,
        Anderson, S. F. \& Margon, B. 1993, ApJ, 403, 610 
\bibitem{ } Angel, J. R. P. \& Stockman, H. S. 1980, ARA\&A, 18, 321 
\bibitem{ } Appenzeller, I., Mandel, H., Krautter, J. et al. 
1995, ApJ, 439, L33 
\bibitem{ } Blandford, R. \& Rees, M. J. 1978 in Pittsburgh Conference
        on BL Lac Objects, ed. A. M. Wolfe 
(Pittsburgh: Univ. Pittsburgh Press),
        p. 328 
\bibitem{ } Brinkmann, W., et al. 1994, A\&A, 288, 433 
\bibitem{ } Bruhweiler, F., Boggess, A., Norman, D. J., Grady, C. A.,
        Urry, C. M., \& Kondo, Y. 1993, ApJ, 409, 199 
\bibitem{ } Celotti, A., Maraschi, L. \& Treves, A. 1991, ApJ, 377, 403 
\bibitem{ } Edelson, R., et al. 1995, ApJ, 438, 120 
\bibitem{ } Falomo, R., Pesce, J. E. \& Treves, A. 1993, ApJ, 411, L63 
\bibitem{ } Kayser, R., Refsdal, S. \& Stabel, R. 1986, A\&A, 166, 36 
\bibitem{ } Krishna, G. \& Subramanian, K. 1991, Nature 349, 766 
\bibitem{ } Maraschi, L., Blades, J. C., Calanchi, C., Tanzi, G. \&
        Treves, A. 1988, ApJ, 333, 660 
\bibitem{ } Morris, S. L., Stocke, J. T., Gioia, I. M., Schild, R. E., Wolter,
        A., Maccacaro, T. \& Della Ceca, R. 1991, ApJ, 380, 49 
\bibitem{ } Quirrenbach, A., et al. 1989, Nature, 337, 442 
\bibitem{ } Schneider, P., Ehlers, J. \& Falco, E. E., in Gravitational Lenses,
        Springer Verlag, Berlin 1992 
\bibitem{ } Schneider, P. \& Weiss, A. 1987, A\&A, 171, 49 
\bibitem{ } Stein, W. A.., O'Dell, S. L. \& Strittmatter, P. A.
        1976, ARA\&A, 14, 173 
\bibitem{ } Stickel, M., Fried, J. W., \& K\"uhr, H. 1993, A\&AS, 98, 393 
\bibitem{ } Stickel, M., Padovani, P., Urry, C. M., Fried, J. W., \& K\"uhr, H.
        1991, ApJ, 374, 431 
\bibitem{ } Stocke, J.T., Wurtz, R.E., \& Perlman, E.S. 1995, ApJ, 454, 55  
\bibitem{ } Urry, C. M., et al. 1993, ApJ, 411, 614 
\bibitem{ } Urry, C. M., \& Mushotzky, R. F. 1982, ApJ, 253, 38 
\bibitem{ } Urry, C. M. \& Padovani, P. 1995, PASP, 107, 803 
\bibitem{ } Urry, C. M., Treves A., Maraschi, L., Marshall, H., Kii, T.
        Madejski, G., Penton, S., Pesce, J. E., Pian, E., 
        Celotti, A., Fujimoto R., Makino F., Otani C., Sambruna R. M.,
        Sasaki K., Shull J.M., Smith P., Takahashi, T., Tashiro M.,
        1997,ApJ , in press 
\bibitem{ } Vermeulen, R.C. \& Cohen, M. H. 1994, ApJ, 430, 467  
\bibitem{ } Vietri, M. \& Ostriker, J.P. 1983, ApJ, 267, 488 
 
\end{thebibliography}
\end{document}